\documentclass[10pt,aps,pra,superscriptaddress,floatfix,nofootinbib]{revtex4-2}
\usepackage{graphicx}
\usepackage{times}
\usepackage{geometry}
\usepackage{rotating}
\usepackage{multirow}
\usepackage{makecell}
\usepackage[utf8]{inputenc}
\usepackage{amssymb}
\usepackage{amsmath}
\usepackage{ragged2e}
\usepackage{footmisc}
\usepackage{comment}
\usepackage{upgreek}
\usepackage{multirow}
\usepackage{booktabs}
\usepackage{datetime}
\usepackage[dvipsnames]{xcolor} 
\definecolor{xlinkcolor}{cmyk}{1,1,0,0}
\usepackage{url}
\usepackage[
 colorlinks=true,    
 linkcolor=xlinkcolor,     
 citecolor=xlinkcolor,     
 filecolor=xlinkcolor,  
 urlcolor=xlinkcolor,      
 final=true
]{hyperref}
\usepackage{enumitem}
\usepackage[left]{lineno}

\setenumerate{itemsep=0mm}
\begin{document}
\title[Environmental $\gamma$-Ray Flux in Hall C at LNGS]{Environmental \texorpdfstring{$\gamma$}--Ray Flux in Hall C at LNGS and Its Correlation with Radon Activity}


\author{L. Luzzi}
\email{lluzzi@ucdavis.edu}
\altaffiliation{Now at \textit{Department of Physics, University of California, Davis, CA 95616, USA}}\affiliation{CIEMAT, Centro de Investigaciones Energéticas, Medioambientales y Tecnológicas, Madrid 28040, Spain}
\author{R. Santorelli}\email{roberto.santorelli@ciemat.es}\affiliation{CIEMAT, Centro de Investigaciones Energéticas, Medioambientales y Tecnológicas, Madrid 28040, Spain}
\author{G. Zuzel}\affiliation{\textit{M. Smoluchowski Institute of Physics, Jagiellonian University, 30-348 Krakow, Poland}}
\author{P. Agnes}\affiliation{\textit{Gran Sasso Science Institute, L’Aquila 67100, Italy}} 
\author{D. Cano-Ott}\affiliation{CIEMAT, Centro de Investigaciones Energéticas, Medioambientales y Tecnológicas, Madrid 28040, Spain}
\author{C.~Ghiano}\affiliation{\textit{INFN Laboratori Nazionali del Gran Sasso, Assergi (AQ) 67100, Italy}}
\author{M.~Laubenstein}\affiliation{\textit{INFN Laboratori Nazionali del Gran Sasso, Assergi (AQ) 67100, Italy}}
\author{T. Mroz}
\thanks{Now at \textit{Henryk Niewodniczanski Institute of Nuclear Physics Polish Academy of Sciences, 31-342 Krakow, Poland}}
\affiliation{\textit{M. Smoluchowski Institute of Physics, Jagiellonian University, 30-348 Krakow, Poland}}
\author{V. Pesudo Fortes}\affiliation{CIEMAT, Centro de Investigaciones Energéticas, Medioambientales y Tecnológicas, Madrid 28040, Spain}
\author{ J. Plaza del Olmo}\affiliation{CIEMAT, Centro de Investigaciones Energéticas, Medioambientales y Tecnológicas, Madrid 28040, Spain}
\author{G.~Vera~Díaz}\affiliation{CIEMAT, Centro de Investigaciones Energéticas, Medioambientales y Tecnológicas, Madrid 28040, Spain}
\begin{abstract}
  
We report a comprehensive measurement of the environmental $\gamma$--ray flux in Hall~C of the Gran Sasso National Laboratory.
A spatial mapping of the radiation was carried out using a high--purity germanium detector mounted on a movable cart and deployed at eight locations within the hall. The detector response function and full--energy--peak efficiencies were determined through \texttt{Geant4} simulations validated with calibrated $\gamma$--ray sources, with particular attention devoted to the efficiency modeling and associated systematic uncertainties. In the energy range of 57–2800 keV, the average $\gamma$--ray flux is measured to be ($0.46 \pm 0.06_{\mathrm{stat}} \pm 0.03_{\mathrm{syst}}$)~cm$^{-2}$~s$^{-1}$. The radon level was monitored for about a month using a radon detector mounted on the same cart, and a clear correlation is observed between the environmental $\gamma$--ray rate and the ambient radon concentration, consistent with the short--lived daughters of $^{222}$Rn. This result represents the first high--precision and efficiency--corrected mapping of the $\gamma$--ray flux in Hall C, substantially improving its radiological characterization and providing key input for future rare--event experiments operating in this hall.
\end{abstract}

\maketitle
\tableofcontents

\section{Introduction}
\label{sec:Intro}

The new generation of rare--event search experiments operating in underground laboratories aims at sensitivities that require an exceptionally detailed understanding of all background contributions, including those associated with environmental $\gamma$--ray radiation in the experimental halls.
A precise and spatially resolved characterization of this component is therefore essential for the design of detector shieldings, the estimation of expected background levels, and, ultimately, the definition of sensitivity goals.
The Gran Sasso National Laboratory (LNGS) is one of the world's largest underground facilities, located beneath 3100 m w.e. of rock overburden.
While some measurements of the environmental $\gamma$--ray flux have been performed in Hall~A and Hall~B of LNGS ~\cite{haffke,malceski,bucci,arpesella}, the information available for Hall~C is extremely limited. 
Hall~C hosts large--scale detectors such as DarkSide--20k (DS--20K)~\cite{ds-20k} and CUPID~\cite{cupid}, whose substantial masses and shielding materials can significantly attenuate or distort the environmental $\gamma$--ray flux, resulting in strong position--dependent variations within the hall.
In addition, Hall~C is characterized by a non--uniform distribution of concrete and rock surfaces, as well as by specific ventilation conditions, which directly affect the spatial and temporal distribution of airborne radon.
These considerations highlight the need for a dedicated and well--controlled measurement campaign.

In this work, we present a measurement of the environmental $\gamma$--ray flux in Hall~C of the LNGS and perform its first systematic spatial mapping. The $\gamma$--ray measurements were carried out using a high--purity germanium (HPGe) detector. To enable a position--dependent study of the environmental background, the HPGe detector (U-type cryostat, liquid nitrogen cooling) and the DAQ system were mounted on a movable cart, allowing the setup to be deployed at multiple locations within Hall~C under identical operating conditions.

In addition, the cart hosted an independent radon monitor, enabling the continuous and simultaneous measurement of the ambient Rn activity alongside the $\gamma$--ray spectra. This configuration was specifically designed to investigate possible correlations between the environmental $\gamma$--ray flux and radon activity inside the experimental hall.

A laptop, also installed on the cart and connected to both the DAQ system and the Rn monitor, was used for remote control of the acquisition, real--time monitoring of detector performance, and data storage.

Particular attention was devoted to the modeling of the HPGe efficiency and to the treatment of systematic uncertainties. The energy--dependent detector efficiency was calculated using a detailed \texttt{Geant4} model implementing the full geometry of the HPGe detector, and subsequently cross--validated over a wide energy range with measurements performed using calibrated $\gamma$--ray sources ($^{133}$Ba, $^{137}$Cs, and $^{60}$Co). After an initial commissioning phase, the detector was deployed at eight locations within Hall~C, performing measurements  of the $\gamma$--ray energy spectrum at each site.

Several studies have reported $\gamma$--ray flux measurements at LNGS (Table~\ref{tab:flux_lngs}). In Hall~A, two measurements performed with HPGe detectors reported fluxes of
$0.23$~cm$^{-2}$ s$^{-1}$ in the energy range 7.4--2734~keV~\cite{malceski} and
$0.73$~cm$^{-2}$ s$^{-1}$ in 0--3000~keV~\cite{bucci}.
A different measurement using a NaI(Tl) detector found a flux of
$0.28 \pm 0.02$~cm$^{-2}$\,s$^{-1}$ in the range 35--3000~keV~\cite{haffke}.
It is worth noting that Ref.~\cite{malceski} derives the flux by counting the detected $\gamma$--rays and normalizing by the detector surface, without applying an efficiency correction, while Ref.~\cite{bucci} does not provide details on the simulation procedure used to convert the measured spectrum into the final flux estimate. Neither work reports an uncertainty on the measured flux. In Hall~B, the flux was measured to be $0.33 \pm 0.02$~cm$^{-2}$\,s$^{-1}$~\cite{haffke} and $0.23$~cm$^{-2}$\,s$^{-1}$~\cite{malceski}. In addition, Refs.~\cite{haffke,malceski} report flux measurements at locations outside the main halls, yielding values of $0.42 \pm 0.06$~cm$^{-2}$\,s$^{-1}$ and $0.38$~cm$^{-2}$\,s$^{-1}$, respectively.
The only flux measurement reported for Hall~C is given in Ref.~\cite{arpesella}, which quotes an approximate value of $1~\mathrm{cm^{-2}\,s^{-1}}$ in the energy range 0--2700~keV. Also in that case, the calculation details are omitted and the result is approximate, without an associated uncertainty.

Table~\ref{tab:flux_lngs} summarizes the $\gamma$--ray flux measurements reported in the literature for various locations at LNGS. The last row lists the flux measured in Hall C reported in this work and represents the first precise and uncertainty--controlled determination of the $\gamma$--ray flux in Hall C. Although the flux measured in this work is of the same order of magnitude as those reported for other LNGS halls, it shows substantial differences in absolute value.

\begin{table*}[htpb]
    \centering
    \footnotesize
    \renewcommand{\arraystretch}{1.1}
    \caption{$\gamma$--ray flux measured at different locations within LNGS as reported in the literature, compared with the result obtained in this work.}
    \begin{tabular}{ccccc}
        \toprule
        \textbf{Ref.}  & \textbf{Location} & \textbf{Detector} & 
        \makecell[c]{\textbf{E Range}\textbf{(keV)}}  & 
        \makecell[c]{\textbf{Flux}\textbf{(cm$^{-2}$ s$^{-1}$)}}   \\
        \midrule
        \multirow{3}{*}{\cite{haffke} (2011)}   
            & Hall A & \multirow{3}{*}{NaI(Tl)} & \multirow{3}{*}{35--3000} & 0.28 $\pm$ 0.02\\
            & Hall B &                        &                           & 0.33 $\pm$ 0.02 \\
            & \makecell[c]{Interferometer\\Tunnel} &                  &  & 0.42 $\pm$ 0.06\\
        \midrule
        \multirow{3}{*}{\cite{malceski} (2013)} 
            & Hall A & \multirow{3}{*}{HPGe} & \multirow{3}{*}{7.4--2734} & 0.23 \\
            & Hall B &                       &                           & 0.23 \\
            & \makecell[c]{Tunnel\\Hall A--Hall B}  & &                & 0.38 \\
        \midrule
        \cite{bucci} (2009) & Hall A & HPGe & 0--3000 & 0.73\\
        \midrule
        \cite{arpesella} (1992) & Hall C & HPGe & 0--2700 & $\sim$1\\
        \midrule
        \textbf{This work} & \textbf{Hall C} & \textbf{HPGe} & \textbf{57--2800} & \textbf{($0.46 \pm 0.06_{\mathrm{stat}} \pm 0.03_{\mathrm{syst}}$)}\\
        \bottomrule
    \end{tabular}
    \label{tab:flux_lngs}
\end{table*}

\section{Experimental setup}
\label{sec:Setup}

\begin{figure*}[htbp]
\centering

\begin{minipage}[t]{0.48\textwidth}
\centering
\raisebox{1cm}{\includegraphics[width=\textwidth]{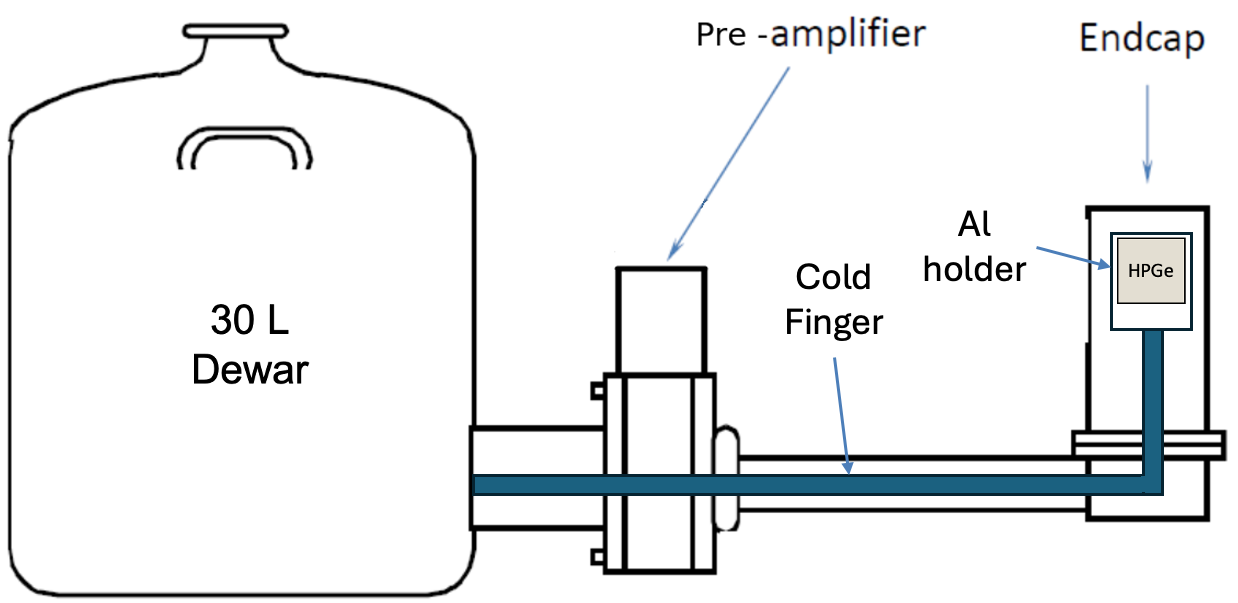}}
\end{minipage}
\hfill
\begin{minipage}[t]{0.48\textwidth}
\centering
\includegraphics[width=\textwidth]{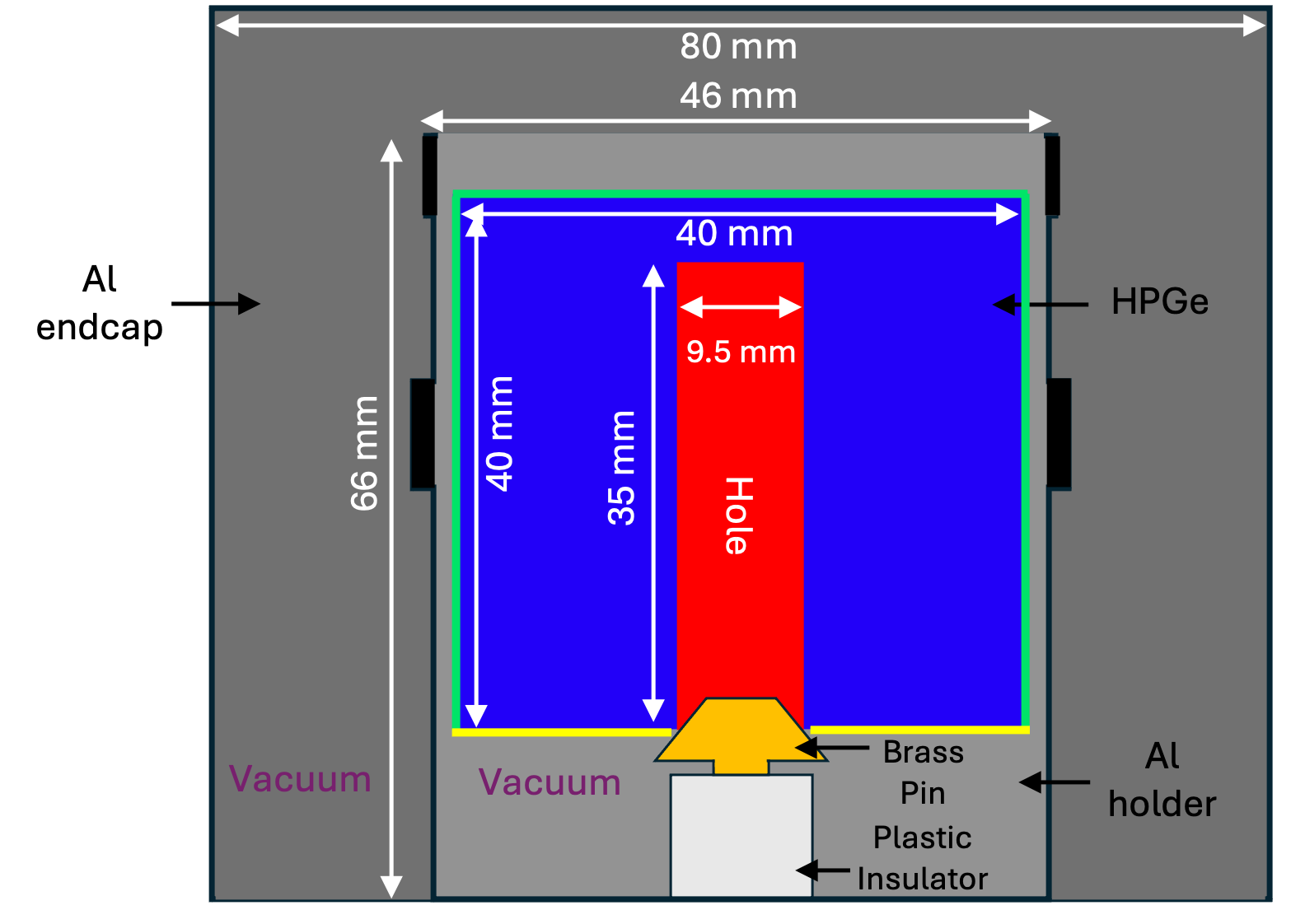}
\end{minipage}

\caption{Left: Scheme of the HPGe detector within the U--type cryostat. 
Right: Two--dimensional schematic of the crystal, aluminum holder, and top section of the end--cap showing the main dimensions. 
The colored edges represent the dead layer included in the Monte Carlo efficiency calculation: top/side dead layer (green) and bottom dead layer (yellow).}

\label{fig:mc_hpge}

\end{figure*}

The HPGe spectrometer used in this work is based on a p--type semi--coaxial detector with 10$\%$ relative efficiency and dimensions of 40 mm in diameter and 40 mm in height. The well has a diameter of 9.5 mm and a depth of 35 mm. The crystal is housed in an aluminum holder with wall thickness varying between 1.0 and 1.5 mm depending on the section. Two reinforcement rings are 3 mm thick. Indium pieces are placed between the crystal and the holder to provide tight mounting. The detector holder is attached to a copper cold finger (with dedicated electrical insulators providing high thermal conductivity) and enclosed within an aluminum end--cap (1 mm thick, 80 mm diameter). The signal is read from the p$^+$~contact (detector well) through the so--called Chinese hat, while the high voltage is applied to the n$^+$~contact (detector holder). Figure~\ref{fig:mc_hpge} shows the details of the spectrometer. The end--cap is significantly larger than the detector holder with the Ge crystal, as the latter comes from a different spectrometer (a hand--held portable spectrometer) and was installed in the available U--type cryostat for the purpose of this study. Since the spectrometer was assembled in--house, all details of the mechanical design (materials and dimensions) are well known, as they were directly measured. This was important for the modeling of the detector response described in the following sections. 

\begin{figure}[htpb]
\centering
\includegraphics[scale=0.43]{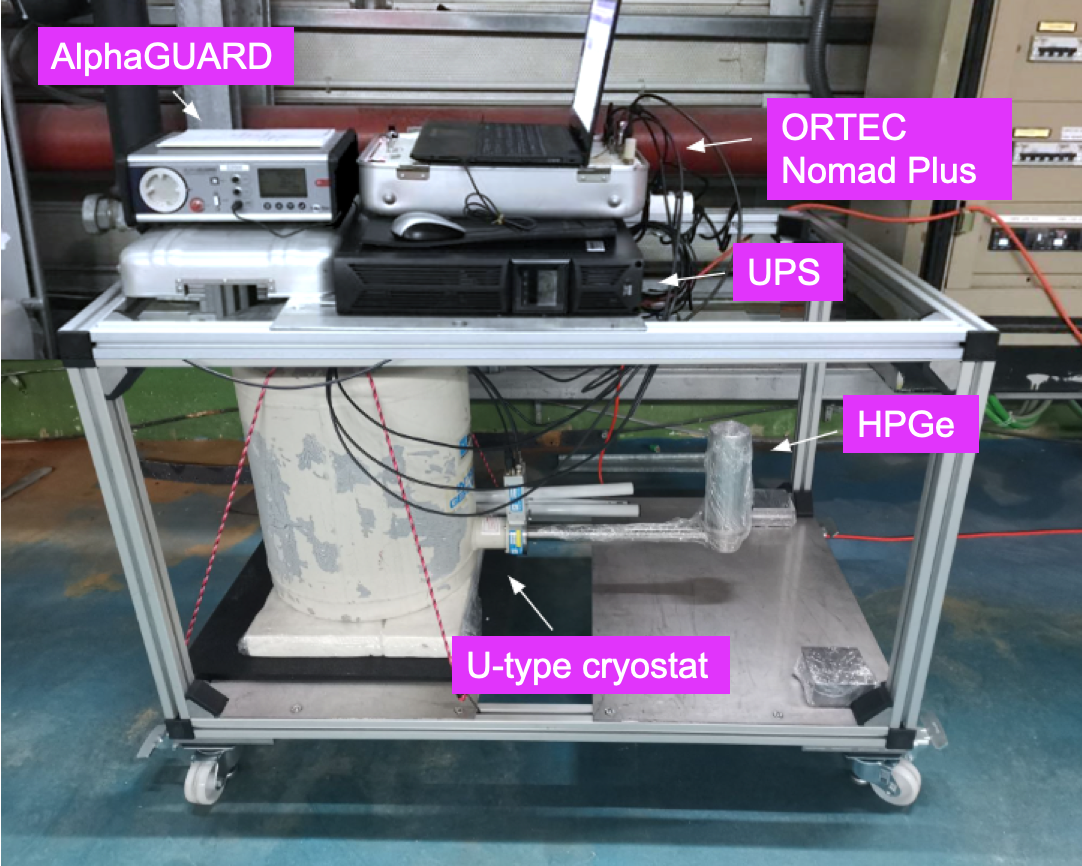}
\caption{Picture of the measurement station, including the HPGe in the U--type cryostat.}
\label{fig:HPGe_LNGS}
\end{figure}

The detector signal is read out using a portable multichannel system (ORTEC Nomad Plus) which interfaces with a laptop. The Nomad Plus provides high--voltage for the detector (+2500 V in our case), the preamplifier power and it is processing of the signal coming from the spectrometer preamplifier~\cite{nomad}. An uninterruptible power supply (UPS) provides automated backup power to both the readout system and laptop, enabling up to 4 hours of operation in case of a mains outage. The entire setup is mounted on a movable cart, allowing the detector to be positioned at multiple locations within Hall~C (Fig. \ref{fig:HPGe_LNGS}). 
Data acquisition is carried out using the ORTEC \textsc{Maestro} multichannel analyzer (MCA) software package.
The trigger configuration results in an energy threshold of 26~keV. A computer macro automatically saves the measured $\gamma$--ray spectrum every hour, allowing us to inspect the temporal stability of the energy scale and to discard individual spectra whenever anomalous conditions were identified (e.g.\ distorted spectral shape or unusually high rates caused by electronic noise or operational activities in the hall).

The $\gamma$--ray measuring station is equipped with a radon monitor AlphaGUARD D2000 detector~\cite{alpha} that incorporates a pulse--counting ionization chamber. It is a portable radon meter with high storage capacity ($\approx$~400 days at a 10--min measuring cycle) and high sensitivity (5~cpm at 100~Bq/m$^{3}$). This instrument is suitable for continuous monitoring of radon activity between 2 and $2\times 10^{6}$~Bq/m$^{3}$. In addition, it also records temperature, pressure and relative humidity, enabling a full characterization of the environmental conditions during the measurements.

\section{Commissioning of the detector}
\label{sec:Commissioning}

\begin{figure}[htbp]
\centering

\includegraphics[width=0.49\textwidth]{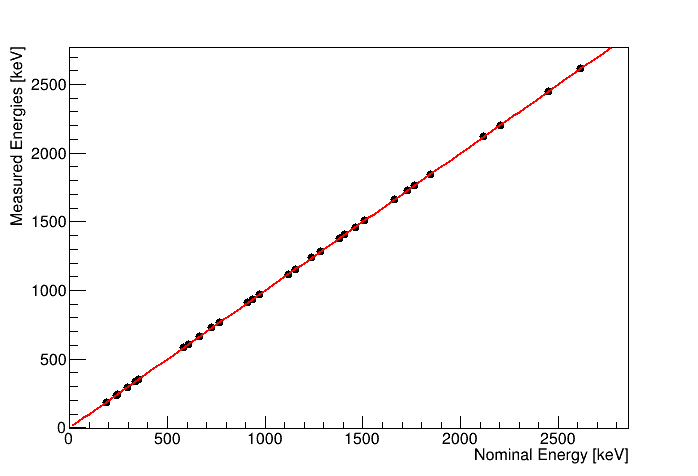}
\hfill
\includegraphics[width=0.49\textwidth]{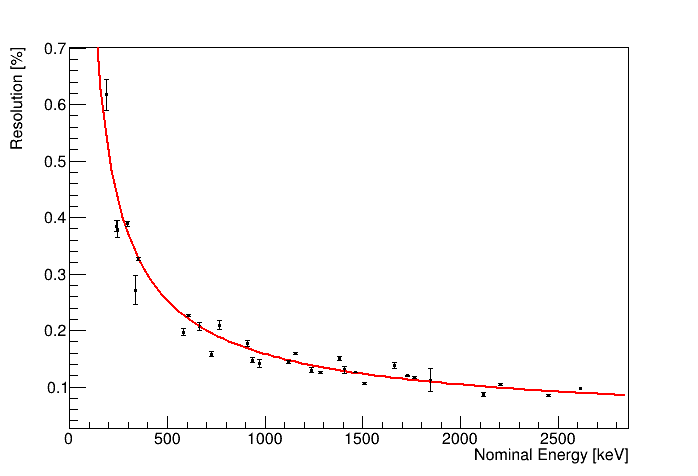}

\caption{Left: Measured peak energies as a function of their nominal values from \cite{radionuclide}. 
Right: Percentage energy resolution (FWHM/E) as a function of the nominal peak energies. The curve has been fitted using the function described by Eq.~(\ref{eq:resolution_fit}).}

\label{fig:res}

\end{figure}

The cart was positioned on the south side of Hall~C, 1 m from the east wall (Fig. \ref{fig:HPGe_LNGS}). Due to the excellent resolution of the germanium detector, the energy scale was assessed directly from the environmental $\gamma$--ray data without the need for an external calibration source.
With a semi--automatic peak--finding algorithm we identified a total of 30 emission lines in the full spectrum. 
By fitting the peaks with a Gaussian plus linear background:
\begin{equation}
    \label{eq:gaus_line}
    f(x)=\frac{1}{\sigma\sqrt{2\pi}}exp \left( \frac{-(x-\mu)^{2}}{2\sigma^{2}} \right) + ax + b \textbf{  },
\end{equation}
we determined the peak centroids in MCA counts with negligible fit uncertainties. The measured peaks' position \cite{radionuclide} were then plotted as a function of the nominal peak energies and fitted with a linear function, as shown in Fig. \ref{fig:res}--left. The resulting calibration function was  
\begin{equation}
    E\;(\text{keV}) = -0.49 + 0.46 \; \text{MCAc} \textbf{  }.
\end{equation}

\begin{figure*}[htpb]
\centering
\includegraphics[scale=0.66]{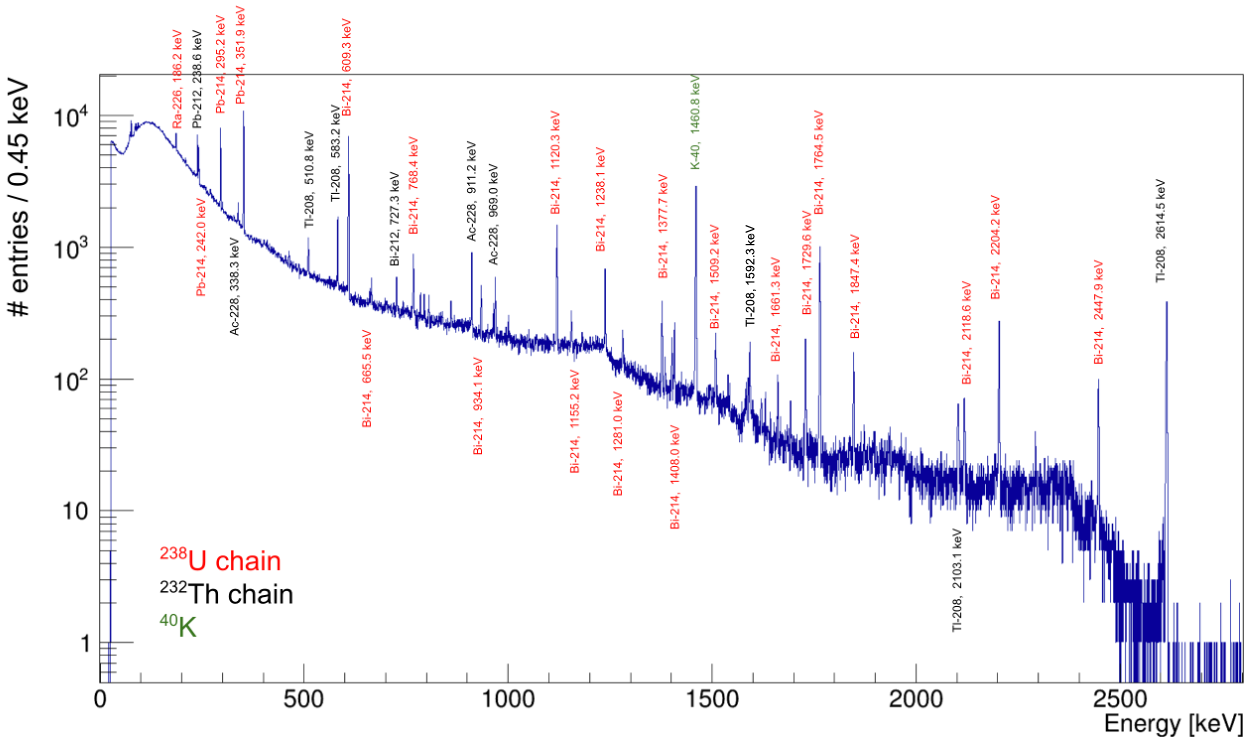}
\caption{Environmental energy--calibrated $\gamma$--ray spectrum with identified peaks from $^{238}$U (red) and $^{232}$Th (black) chains and from $^{40}$K (green).}
\label{fig:gamma_spectrum_all_peaks}
\end{figure*}

The detector response was linear over the full energy range considered. Figure~\ref{fig:gamma_spectrum_all_peaks} shows the energy--calibrated environmental 
$\gamma$--ray spectrum, with the identified emission lines from the $^{238}$U (red) and 
$^{232}$Th (black) decay chains, as well as from $^{40}$K (green). 
The spectrum corresponds to 
a total acquisition time of 76 hours. During this period, an automated acquisition job in 
ORTEC \textsc{Maestro} was used to segment the data into consecutive 1--hour files. This 
approach allowed for continuous monitoring of the detector stability and for the identification 
of any periods potentially affected by abnormal operating conditions—such as deviations in the 
trigger rate or distortions in the spectral shape. No such anomalies were observed, and thus the entire dataset was retained for the analysis. The performance of the spectrometer, including the response of the data acquisition system, was verified during commissioning using test pulses injected through the preamplifier test input. No evidence of signal losses or trigger inefficiencies was observed in the energy range relevant for this work. The dead time reported by the MCA electronics through the \textsc{Maestro} software was monitored for each acquisition and was found to remain below 0.1\% throughout the measurement campaign. Its effect on the measured count rate is therefore negligible compared to the other sources of uncertainty.

The region of interest (ROI) is defined as 57--2800 keV, with the lower bound set by the electronic noise threshold corresponding to the local minimum observed in the low--energy region, and the upper bound chosen to include the residual continuum and pile--up events extending beyond the $^{208}$Tl line at 2614.5~keV.

For each 1--hour file, the $\gamma$--ray rate within the ROI was computed by dividing the 
integral of the spectrum by the corresponding live time. 
The overall rate was then 
obtained as the average of the rates over all hourly files, yielding a total value of 
$(15.00 \pm 0.11)$~s$^{-1}$. The quoted uncertainty represents the root--mean--square (RMS) variation of the hourly $\gamma$--ray rates.

The energy resolution of the HPGe detector was characterized as a function of the $\gamma$--ray energy by analyzing the relative width of the full--energy peaks. 
For each identified $\gamma$--ray line, the peak width was extracted from a Gaussian plus linear fit and expressed in terms of the relative full width at half maximum,
\begin{equation}
\label{eq:resolution_def}
R(E) = \frac{\mathrm{FWHM}(E)}{E},
\end{equation}
where $\mathrm{FWHM}(E)$ denotes the full width at half maximum of the peak at energy $E$.
Following the standard description of the energy resolution in semiconductor detectors~\cite{Knoll}, the energy dependence of the peak width was parameterized as
\begin{equation}
\label{eq:FWHM}
\mathrm{FWHM}^2(E)=a+bE+cE^2,
\end{equation}
where $a$, $b$, and $c$ are free parameters accounting for electronic noise, statistical fluctuations in charge collection, and higher--order contributions.

The measured relative energy resolution as a function of the peak energy is shown in Fig. \ref{fig:res}--right and is fitted using the function
\begin{equation}
\label{eq:resolution_fit}
f(E) = \frac{\sqrt{a + bE + cE^2}}{E}.
\end{equation}
The fit yields $a = 0.68 \pm 0.02$ keV$^{2}$, $b = 1.83 \pm 0.02\times10^{-3}$ keV, while the coefficient of the quadratic term is consistent with zero within the fit uncertainty. The coefficient $b$ corresponds to a Fano factor of $F=0.11\pm0.01$, in good agreement with values typically observed for HPGe detectors~\cite{lutz}. A FWHM of 1.9 keV was measured at the 1332.5 keV $\gamma$--ray line of $^{60}$Co, in agreement with typical values for this type of HPGe detectors~\cite{Knoll}.
The resulting curve quantitatively characterizes the detector performance, with the relative energy resolution improving from about 0.6\% at $\sim$100~keV to below 0.1\% above 2~MeV, demonstrating a reliable detector response over the energy range relevant for the environmental $\gamma$--ray measurements and well suited to the purposes of this study. Small deviations from the fitted curve are observed for a few low-energy $\gamma$-ray lines (below about 300 keV), where the determination of the peak width is more sensitive to the local background shape and nearby spectral structures.

We measured the intrinsic background of the HPGe in a dedicated run. 
\begin{figure}[htpb!]
    \centering
    \includegraphics[width=.6\columnwidth]{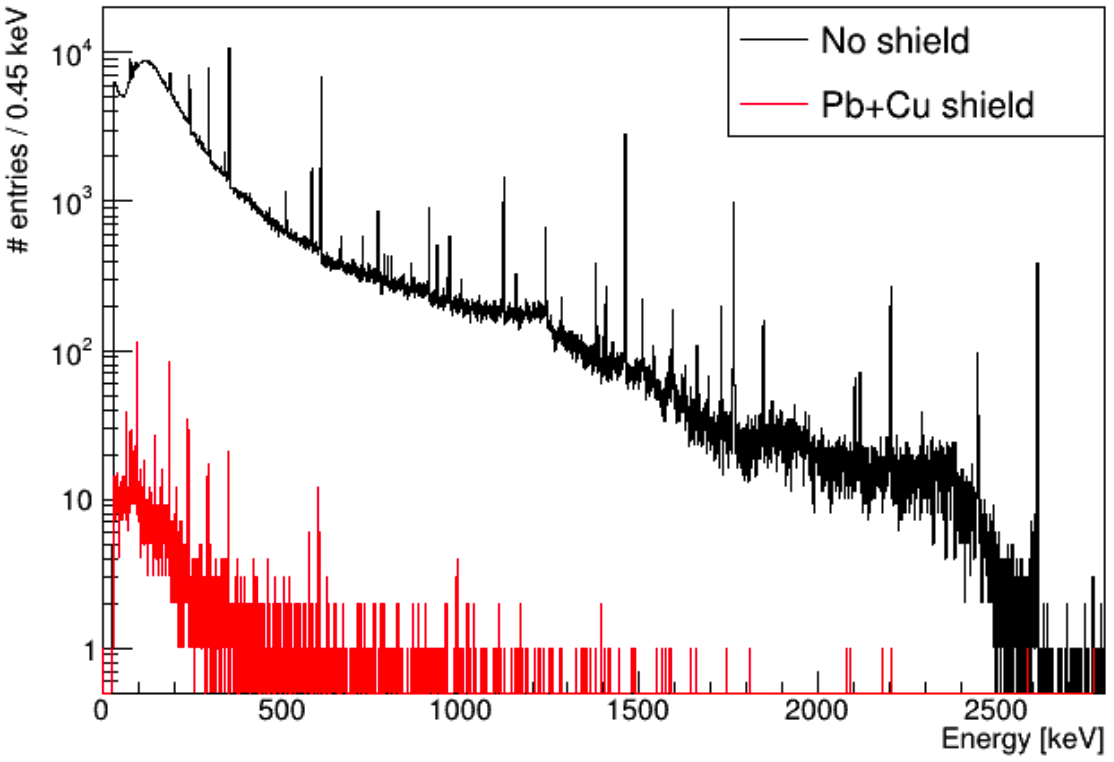}
    \caption{In black, the $\gamma$--ray spectrum from Hall C walls; in red, the intrinsic background
of the HPGe measured with a Pb + Cu shield surrounding the detector.}
    \label{fig:bkg}
\end{figure}A 10 cm thick lead (Pb) castle was installed around the endcap, covering the full 4$\pi$ solid angle, with an additional 10 cm copper (Cu) liner placed inside to suppress $\gamma$–rays produced by Bremsstrahlung from the $\beta$–decay of $^{210}$Bi in the lead. The resulting background spectrum, shown in Fig.~\ref{fig:bkg}, is compared to the environmental $\gamma$–ray spectrum from the Hall C walls; both are normalized to the same live time. 
In the shielded configuration, X--ray lines from $^{210}$Bi become clearly visible at low energy.
The intrinsic background rate of the detector was found to be $(0.080 \pm 0.004)$ s$^{-1}$, representing less than 1\% of the unshielded $\gamma$--ray rate. For this reason, it can be considered negligible for the purposes of the measurement presented in this work. Moreover, this value is conservative, since the measured rate includes not only the intrinsic $\gamma$--ray contributions from the Pb and Cu shielding materials themselves, but also the contribution from radon daughters in the air inside the shield, which was not airtight and therefore continuously exchanged air with the surroundings.
\section{Data taking in Hall C}
\label{sec:Data_taking}

\begin{figure}[htpb]
\centering
\includegraphics[scale=0.49]{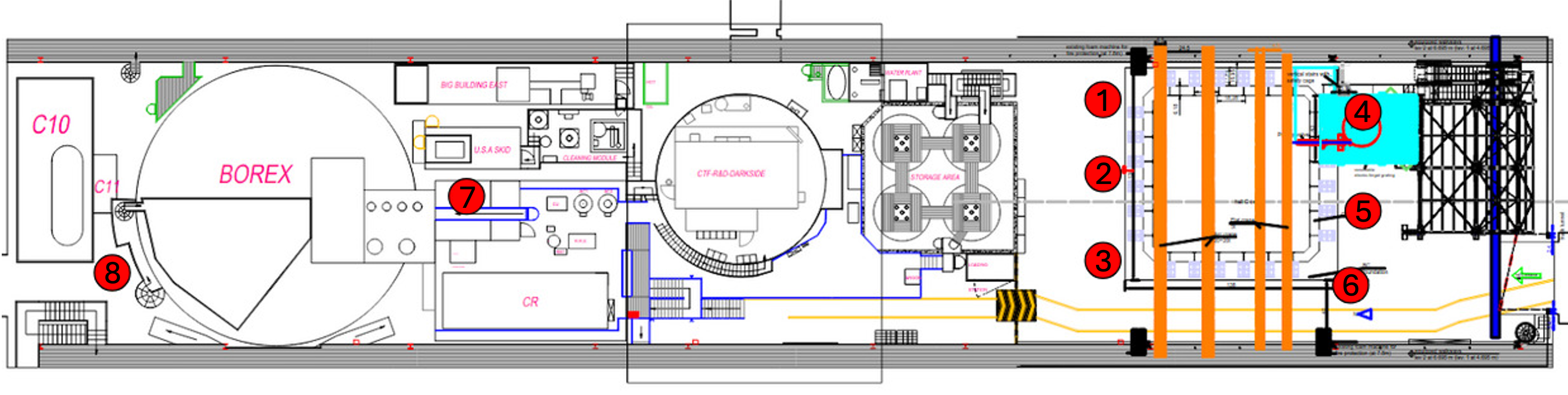}
\caption{Plan view of Hall C showing the eight measurement locations.}
\label{fig:hallC_ds}
\end{figure}

Between 12 and 20 February 2025, the HPGe detector was deployed at eight locations within Hall~C to map the spatial variations of the environmental $\gamma$--ray flux. Six positions (1--6) were located along the southern side of the hall, surrounding the DS--20k cryostat; one position (7) was placed on a scaffolding platform above the Borexino experiment; and one position (8) was located on the northern side, in front of the CUPID experiment. Figure~\ref{fig:hallC_ds} shows a plan view of Hall~C with the measurement locations indicated.

All measurements were carried out under identical hardware and DAQ conditions, as described in Sec.~\ref{sec:Setup} and~\ref{sec:Commissioning}. The live time of each measurement ranged between 6 and 19~hours. Table~\ref{tab:rates_ds_tab} reports the average $\gamma$--ray rates measured at the eight positions within the ROI, together with the overall average rate for the entire hall.

\begin{table}[htbp]
    \centering
    \renewcommand{\arraystretch}{1.3}
    \footnotesize
    \caption{$\gamma$--ray rates measured at the eight positions within Hall~C in the energy range 57--2800 keV.}
    \begin{tabular}{ccc}
        \toprule
        \textbf{Position} & \textbf{Live Time (h)}  & \textbf{Rate (s$^{-1}$)} \\
        \midrule
        1 & 19 & 14.59 $\pm$ 0.09  \\
        2 & 16 & 14.80 $\pm$ 0.76  \\
        3 & 16 & 15.71 $\pm$ 0.53  \\
        4 & 6  & 15.11 $\pm$ 0.20  \\
        5 & 16 & 15.74 $\pm$ 0.11  \\
        6 & 7  & 14.66 $\pm$ 0.11  \\
        7 & 16 & 19.27 $\pm$ 0.13  \\
        8 & 16 & 18.33 $\pm$ 0.10  \\
        \bottomrule
    \end{tabular}
    \label{tab:rates_ds_tab}
\end{table}
\noindent Positions 7 and 8 exhibit rates approximately 28\% and 21\%, respectively, higher than those measured at positions 1--6. 
This difference is plausibly associated with the presence of massive structures inside the hall, which can modify the $\gamma$--ray flux from the surrounding rock or affect the airborne radon activity by impacting the air recirculation induced by the ventilation system. The overall average rate over the eight positions is (16.03 $\pm$ 1.79) s$^{-1}$, where the associated uncertainty is given by the standard deviation of the measurements across the different positions, reflecting the observed spatial variability. 

The $\gamma$--ray flux is derived from the measured count rate by correcting for the detector efficiency and the geometrical acceptance. For a given analysis energy threshold, the flux through a surface area $S$ is defined as
\begin{equation}
\Phi(E) = \frac{R(E)}{\varepsilon(E) S}  ~~ ,
\end{equation}
where $R(E)$ is the measured $\gamma$--ray rate above the selected energy threshold and $\varepsilon(E)$ is the corresponding energy--dependent detection efficiency, evaluated using dedicated Monte Carlo simulations.
The procedure, described in detail in the following section, is divided into two steps: first, the validation of the detector geometry implemented in the Monte Carlo model; second, the simulation of the environmental $\gamma$--ray flux in Hall~C.
\section{Monte Carlo modeling of the detector response}
\label{sec:MC}

\begin{figure}[htpb]
    \centering
    \includegraphics[width=0.45\columnwidth]{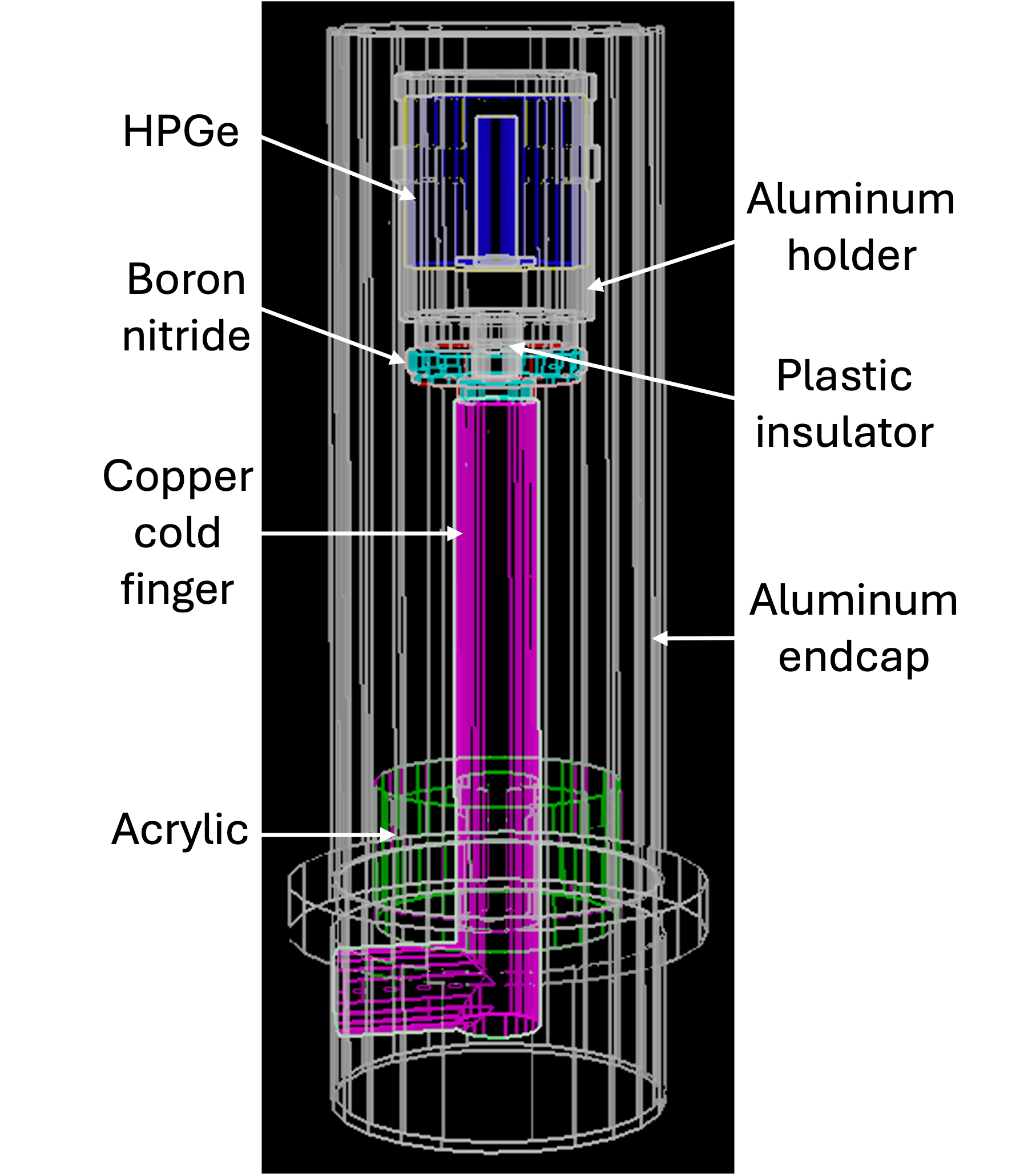}
    \caption{Geometry of the HPGe spectrometers implemented in \texttt{Geant4}.}
    \label{fig:mc_all}
\end{figure}

The HPGe detector geometry was modeled in \texttt{Geant4} \cite{geant}, as shown in Fig.~\ref{fig:mc_hpge}--right. It includes the germanium crystal (semi--coaxial with the p$^+$ contact in the bore hole), its aluminum holder (with reinforcing rings), the aluminum end-cap, a plastic insulator (PTFE) providing electrical isolation between the signal read-out pin (the so--called Chinese hat, brass) and the holder, an acrylic element that supports the detector holder assembly, and a boron--nitride discs that act as thermal conductors and electric insulators for the detector holder. Figure~\ref{fig:mc_all} shows details of the spectrometer geometry implemented in the Monte Carlo model.

A precise determination of the environmental $\gamma$--ray flux requires an accurate calculation of the detector efficiency over a broad energy range. Such a calculation depends critically on a detailed and realistic description of the detector geometry, including passive regions and surrounding materials that significantly affect the detector response, in particular at low energies. For this reason, the Monte Carlo simulation is developed in two successive steps. In the first step, the detector geometry implemented in the simulation is validated using calibration measurements performed with well--characterized $\gamma$--ray sources placed at known positions around the detector. This validation ensures that the simulated response accurately reproduces the measured full--energy--peak efficiencies and provides a reliable description of the detector behavior. 

In the second step, the validated detector model is used to simulate the response to external environmental $\gamma$--rays, allowing the calculation of the energy--dependent detection efficiency required for the flux determination.

\subsection{Validation of the Monte Carlo–simulated crystal geometry using calibrated sources}

To validate the MC geometry, we acquired calibration data using three radioactive sources: $^{133}$Ba (80.9 keV, 276.4 keV, 302.9 keV, 356.9 keV, 383.8 keV), $^{137}$Cs (661.6 keV), and $^{60}$Co (1173.2 keV, 1332.0 keV). 

The $^{137}$Cs source was deployed at six well--defined positions around the detector: four lateral positions (north, south, west, and east), as well as a top and a bottom position, all located 50~cm from the endcap. Among the lateral positions, the east position corresponds to the side of the LN$_{2}$ dewar, whereas the north, south, and west positions lie along the remaining azimuthal directions. 
Data with $^{133}$Ba and $^{60}$Co were acquired only from the top position. To isolate the contribution from the source alone, the spectrum acquired with the source is obtained by subtracting the ambient $\gamma$--ray spectrum measured without the source.

For each source position, a corresponding MC simulation was performed by generating $10^{8}$ $\gamma$--rays from point--like sources placed at the same location around the detector.
The goal of this validation procedure is to ensure agreement between the measured and simulated full--energy--peak efficiencies (FEPEs) for all $\gamma$--ray lines and configurations. 
For a line energy $E_{\gamma}$, the data and MC FEPEs are defined as
\begin{equation}
\varepsilon^{\text{data}}(E_{\gamma}) = \frac{N^{\text{data}}_{\text{FEP}}(E_{\gamma})}{A\, t \, BR(E_{\gamma})},
\quad
\varepsilon^{\text{MC}} (E_{\gamma})   = \frac{N^{\text{MC}}_{\text{FEP}}(E_{\gamma})}{N_{\text{sim}} \, BR(E_{\gamma})},
\label{eq:FEP}
\end{equation}
where $N_{\text{FEP}}$ is the number of events in the full--energy peak (FEP), $A$ is the nominal source activity, $t$ is the live time, $N_{\rm sim}$ is the total number of simulated decays, and $BR(E_{\gamma})$ is the branching ratio of the line.

A typical p--type coaxial HPGe detector has a Li--drifted n$^+$contact on the outer surface of the crystal. Its typical thickness is about 0.7 mm (for new detectors). This is an inactive volume with incomplete charge collection, leading to attenuation of low--energy $\gamma$--rays, with effects becoming particularly relevant below $\sim$100~keV~\cite{lee}. For our detector, which is about 30 years old, thickness of this dead layer is not precisely known (it can be much thicker than 0.7 mm due to residual Li diffusion in Ge). For this reason, the dead layer is treated as an effective correction parameter and parameterized in terms of two independent components with different thicknesses: a top/side layer and a bottom layer, going up to the groove (Fig.~\ref{fig:mc_hpge}--right)~\cite{jesko}. As mentioned earlier, other details of the spectrometer are rather well known. The reason for considering the top/side and the bottom dead layers separately is that, during the detector fabrication process, when Li is evaporated onto Ge and diffuses into the bulk, these two regions are formed in two separate steps: first the top/side is formed, then the bottom~\cite{jany}.

To determine the thickness of each dead layer component, we compare the measured FEPEs with the corresponding MC predictions for $^{133}$Ba and $^{137}$Cs. The optimization identifies the combination of the two dead layer thicknesses $(t_{\rm top/side}, t_{\rm bottom})$ that best reproduces the 81.9 and 356.9~keV lines of $^{133}$Ba and the 661.6~keV line of $^{137}$Cs in the top, bottom, and north positions.

For each simulated event, all single energy--deposit interactions in the crystal’s active region are summed to obtain the total deposited energy.
The resulting energy--deposition spectrum is histogrammed, and the MC FEPE for a $\gamma$--ray line of energy $E_{\gamma}$ is then computed following Eq.~(\ref{eq:FEP}).

\begin{figure}[htbp]
    \centering
    \includegraphics[width=.6\columnwidth]{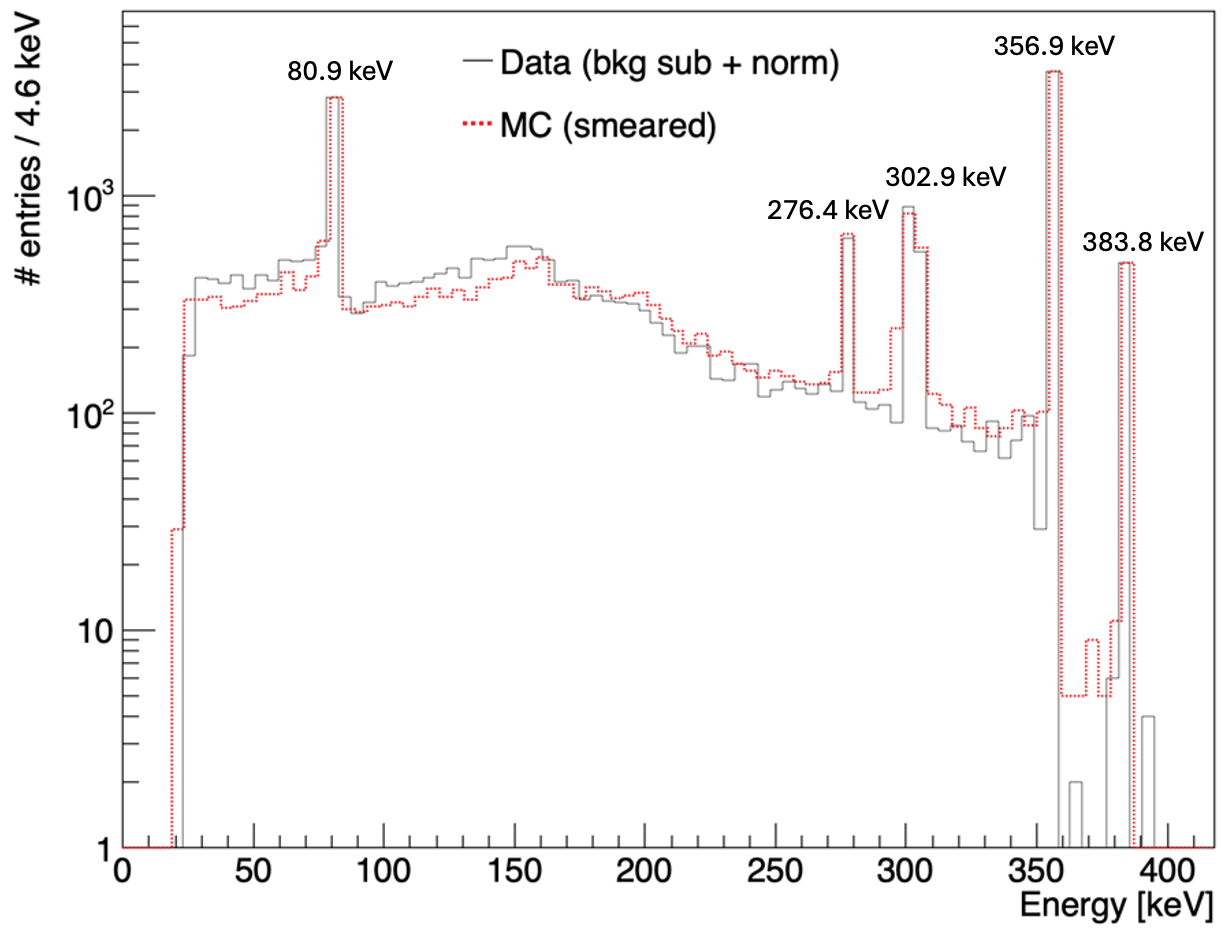}
    \caption{Data--MC comparison of the $^{133}$Ba spectrum for the dead layers combination $(t_{\rm top/side}^{\rm best},\, t_{\rm bottom}^{\rm best}) = (1.7, 1.7)$ mm.}
    \label{fig:ba_mc_data}
\end{figure}

An algorithm varies the two layer thicknesses from 0 to 3.5 mm in steps of 0.1 mm to find the combination that minimizes the difference between $\varepsilon^{\text{data}}$ and $\varepsilon^{\text{MC}}$.
This scan yields the best--fit values $(t_{\rm top/side}^{\rm best}, t_{\rm bottom}^{\rm best}) = (1.7, 1.7)~\mathrm{mm}$.
Figure~\ref{fig:ba_mc_data} shows the data--MC comparison with the $^{133}$Ba source placed at 50 cm above the endcap. 
The data spectrum is obtained by subtracting the spectrum acquired without the source (background) and normalizing its live time to match the MC. All FEPs agree in terms of event population. The Compton region between 26~keV and 150~keV is more populated in the data, likely due to $\gamma$--ray backscattering on surrounding materials that are not included in the simulation. This difference is negligible for the purposes of this analysis, since the detection efficiency is derived from full--energy peaks, which are correctly reproduced by the simulation.

With the best--fit values $(t_{\rm top/side}^{\rm best},\, t_{\rm bottom}^{\rm best})$, we then compute the data and MC FEPEs for all calibration lines and all source positions, and evaluate their ratio $\varepsilon^{\text{data}}$/$\varepsilon^{\text{MC}}$, as reported in Table \ref{tab:eff}.  

\begin{table*}[htbp]
    \centering
    \footnotesize
    \renewcommand{\arraystretch}{1.1}
    \caption{FEPEs comparison between data and Monte Carlo for the three calibration sources.}
    \begin{tabular}{cccccccc}
        \toprule
        \textbf{Source} & \textbf{E (keV)} & \textbf{A (kBq)} & \textbf{BR (\%)} & \textbf{Position around HPGe} &    \textbf{$\boldsymbol{\varepsilon^{\text{data}}}(\times10^{-5}$)} &    \textbf{$\boldsymbol{\varepsilon^{\text{MC}}}(\times10^{-5}$)} &    \textbf{$\boldsymbol{\varepsilon^{\text{data}}/\varepsilon^{\text{MC}}}$}\\
        \midrule
        \multirow{5}{*}{$^{133}$Ba} &
        80.9 & \multirow{5}{*}{48.5} & 33.3 & \multirow{5}{*}{Top} & 8.48 $\pm$ 0.27 & 8.12 $\pm$ 0.16 & 1.04 $\pm$ 0.04\\
        & 276.4 & & 7.1 & & 8.88 $\pm$ 0.33 & 9.63 $\pm$ 0.37 & 0.92 $\pm$ 0.05\\
        & 302.9 & & 18.3 & & 7.82 $\pm$ 0.26 & 8.00 $\pm$ 0.21 & 0.98 $\pm$ 0.04 \\
        & 356.9 & & 62.0 & & 5.95 $\pm$ 0.19 & 5.97 $\pm$ 0.10 & 1.00 $\pm$ 0.04\\
        & 383.8 & & 8.9 & & 5.53 $\pm$ 0.22 & 5.53 $\pm$ 0.25 & 1.01 $\pm$ 0.06\\
        \midrule
        \multirow{6}{*}{$^{137}$Cs} 
      & \multirow{6}{*}{661.6} & 
      \multirow{6}{*}{169.5} & \multirow{6}{*}{85.0} & North & 4.57 $\pm$ 0.14 & 4.81 $\pm$ 0.07 & 0.95 $\pm$ 0.03\\
         &  &  & & South & 4.55 $\pm$ 0.14 & 4.90 $\pm$ 0.07 & 0.93 $\pm$ 0.03\\
         &  &  & & East & 3.72 $\pm$ 0.12 & 4.37 $\pm$ 0.06 & 0.85 $\pm$ 0.03\\
         &  &  & & West & 4.63 $\pm$ 0.14 & 4.66 $\pm$ 0.07 & 0.99 $\pm$ 0.03\\
         &  &  & & Top & 3.32 $\pm$ 0.11 & 3.33 $\pm$ 0.06 & 0.99 $\pm$ 0.04\\
         &  &  & & Bottom  & 0.61 $\pm$ 0.02 & 0.61 $\pm$ 0.03 & 1.01 $\pm$ 0.06\\
        \midrule
        \multirow{2}{*}{$^{60}$Co} &
        1173.2 & \multirow{2}{*}{4.8} & 99.9 & \multirow{2}{*}{Top} & 2.07 $\pm$ 0.10 & 2.11 $\pm$ 0.05 & 0.98 $\pm$ 0.05\\
        & 1332.0 & & 99.9 & & 1.94 $\pm$ 0.09 & 1.87 $\pm$ 0.04 & 1.04 $\pm$ 0.06 \\
        \bottomrule
    \end{tabular}
    \label{tab:eff}
\end{table*}

The FEPE measured for $^{137}$Cs in the top position is lower than in the lateral positions because the source (as well as the $^{133}$Ba and $^{60}$Co sources) was placed on a wooden support.  The same support is included in the MC model, resulting in agreement between the measured and simulated efficiencies for this configuration.

The $^{137}$Cs FEPE in the bottom position is the smallest among all configurations. In this case, the source lies beneath a 5~mm stainless--steel plate (see Fig.~\ref{fig:mc_hpge}--left) and below the entire endcap assembly, including the copper cold finger.  
Both structures are modeled in the simulation, and the predicted efficiencies are consistent with the measurements.

In contrast, the $^{137}$Cs FEPE in the east position is approximately 15\% lower in data than in simulation.  
This discrepancy is plausibly explained by the presence of additional material between the source and the crystal — namely the preamplifier block and two BNC connectors with cables (Fig.~\ref{fig:HPGe_LNGS}) — whose exact geometry is difficult to reproduce in detail.
These components introduce extra attenuation, reducing the measured FEPE relative to the MC prediction.

\subsection{Monte Carlo modeling of the HPGe efficiency}
\label{sec:MC_ext_gammas}

Once the MC geometry and the optimal combination of dead layer thicknesses were validated using calibration sources, we simulated the detector response to external $\gamma$--rays to determine the energy--dependent detection efficiency required for the flux calculation.

\begin{figure}[htpb]
\centering
\includegraphics[scale=0.38]{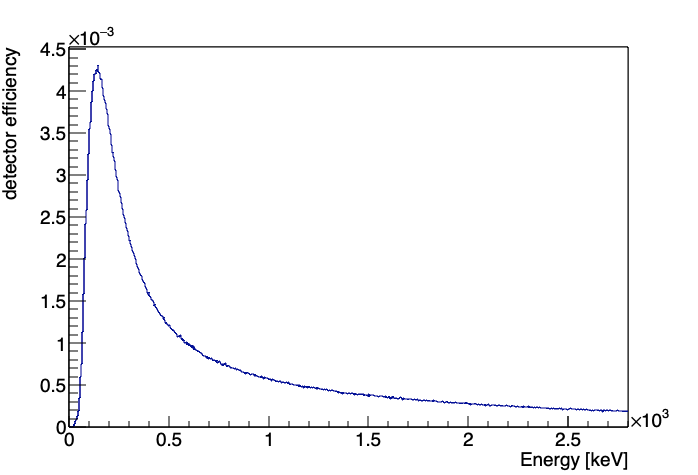}
\caption{Detector efficiency as a function of the initial $\gamma$--ray energy, obtained from MC simulations.}
\label{fig:eff_curve}
\end{figure}

Environmental $\gamma$--rays were generated from the surface of a spherical source with a radius of 23\,cm centered on the detector. A total of $10^{10}$ events were simulated, with initial energies uniformly distributed between 0 and 2800\,keV. Emission points were sampled uniformly over the sphere surface, and initial directions followed a cosine distribution pointing toward the detector, ensuring a uniform incident flux through the spherical boundary~\cite{john}.
The detector efficiency as a function of the initial $\gamma$--ray energy (Fig.~\ref{fig:eff_curve}) is obtained by dividing the number of full--energy deposition events in the Ge crystal by the generated spectrum. 
The efficiency peaks at 0.43\% for initial energies between 135 and 145\,keV, and then gradually decreases with increasing energy. The average efficiency over the ROI is 0.074\%. At low energies, the dead layer leads to a strong suppression of the detector response: the efficiency drops below 0.08\% for $E < 60\,\mathrm{keV}$ and falls further to below 0.006\% for $E < 30\,\mathrm{keV}$.

\section{ \texorpdfstring{$\gamma$}{gamma}--ray flux calculation}
\label{sec:Flux_calc}

The $\gamma$--ray energy--dependent flux at a given detector position is obtained by converting the measured energy spectrum at that location into a flux spectrum. For each energy bin $i$, the flux $\Phi_i$ is calculated by dividing the measured number of counts $C_i$ by the corresponding detector efficiency $\varepsilon_i$ (Fig.~\ref{fig:eff_curve}), the spherical source surface area $S$, and the live time $t$ of the corresponding run:
\begin{equation}
\label{eq:flux}
\Phi_i = \frac{C_i}{\varepsilon_i , S  t} \textbf{ }.
\end{equation}
The total flux over the full energy range is obtained as $\sum_{i=1}^{N} \Phi_{i}$, where $N$ is the number of energy bins. The statistical uncertainty arises from Poisson counting statistics in the measured spectrum and from the MC statistical uncertainty in the efficiency calculation. 
Propagating these uncertainties through Eq.~(\ref{eq:flux}) yields
\begin{equation}
     \Delta \Phi_{stat} = \frac{1}{S\text{   }t}  \sqrt{ \sum_{i=1}^{N} \left[ \left( \frac{\Delta C_{i}}{\varepsilon_{i}} \right)^2 + \left( -\frac{ C_{i} \Delta \varepsilon_{i}}{\varepsilon_{i}^{2}} \right)^2 \right] } \text{    },
\end{equation}
where $\Delta C_{i}$ is the statistical uncertainty on the counts in bin $i$ and $\Delta \varepsilon_{i}$ is the statistical uncertainty on the corresponding efficiency.

The systematic uncertainty associated with the detector’s geometrical efficiency was estimated from the data/MC efficiency ratios measured with the $^{137}$Cs source at different positions around the detector (Table \ref{tab:eff}). The $^{137}$Cs source is the only calibration source that was deployed at multiple locations, and therefore it is the only one that provides direct information on the position--dependent response of the detector.
The percentage uncertainty of the efficiency ratios is
\[
\frac{\sigma_{\bar{\rho}}}{\bar{\rho}}
    = \frac{0.058}{0.953}
    = 0.061,
\]
where $\rho = \varepsilon^{\text{data}}_{\text{Cs}}/\varepsilon^{\text{MC}}_{\text{Cs}}$ for a given source position,
$\sigma_{\bar{\rho}}$ is the RMS of the $\rho$ distribution, and 
$\bar{\rho}$ is its average value across the six source positions.
This quantity represents the systematic uncertainty associated with the
geometrical efficiency of the detector.
The corresponding uncertainty on the measured flux is then obtained as
\begin{equation}
\Delta\Phi_{\text{syst}}
= \Phi\,\frac{\sigma_{\bar{\rho}}}{\bar{\rho}} \; .
\end{equation}

\begin{figure}[b!]
\centering
\includegraphics[scale=0.37]{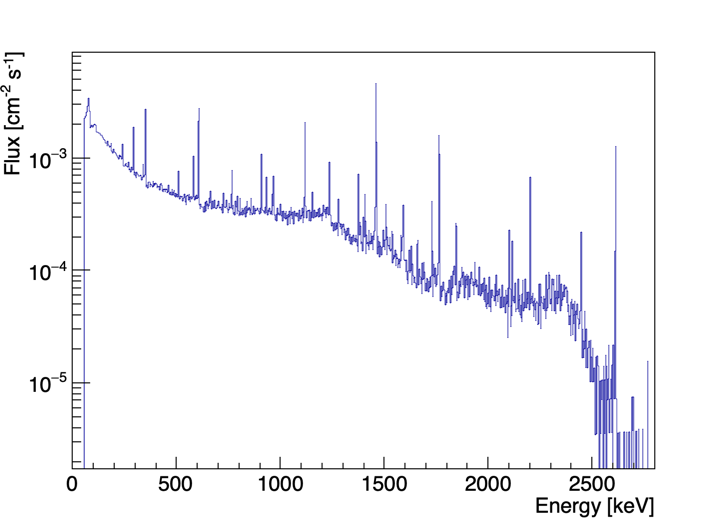}
\caption{$\gamma$--ray flux energy spectrum measured with the detector at position~1 within the ROI.}
\label{fig:flux_pos1}
\end{figure}

Figure~\ref{fig:flux_pos1} shows the $\gamma$--ray flux spectrum in the ROI measured with the detector at position~1, representative of the spectra acquired at the different locations in the hall. 

The $\gamma$--ray fluxes measured across Hall C (Table~\ref{tab:flux_ds_tab}) are broadly consistent with each other, showing moderate position--dependent variations. However, measurements at positions~1–6 reflect similar environmental conditions and yield flux values whose average is approximately 32\% lower than those measured at positions 7 and 8. The average $\gamma$--ray flux is $(0.46 \pm 0.06_{\mathrm{stat}} \pm 0.03_{\mathrm{syst}})$~cm$^{-2}$~s$^{-1}$, where the statistical uncertainty is given by the standard deviation of the measurements across the eight positions, accounting for the observed spatial variability, while the systematic uncertainty corresponds to the average of the individual systematic uncertainties.


The average radon activity measured during the data taking at each position is also reported in the last column of Table~\ref{tab:flux_ds_tab}, showing that the higher $\gamma$--ray fluxes correspond to higher Rn activities in air. This observation suggests a possible impact of the ambient Rn activity on the measured environmental $\gamma$--ray flux, motivating a preliminary investigation of their correlation. 

The average fluxes of the most prominent $\gamma$-ray lines identified in the measured spectra are reported in Table~\ref{tab:flux_ds_tab_isotopes}. 
\begin{table*}[htpb]
    \centering  
    \renewcommand{\arraystretch}{1.3}
    \footnotesize
    \caption{$\gamma$--ray rate and fluxes measured at the eight positions within Hall C around DS--20k, in the energy range 57--2800 keV. The radon activity measured at each position during the same live time as the $\gamma$--rays is also shown. The first and second uncertainties on the flux represent the statistical and systematic contributions, respectively, whereas the quoted rate and radon activity include only the statistical uncertainty.}
    \begin{tabular}{ccccc}
        \toprule
        \textbf{Position} & \textbf{Live Time (h)}  & \textbf{Rate (s$^{-1}$)}  & \textbf{Rn Activity (Bq m$^{-3}$)} & \textbf{Flux ( cm$^{-2}$ s$^{-1}$)} \\
        \midrule
        1  & 19 & 14.59 $\pm$ 0.09  & 82.8 $\pm$ 12.7 & 0.41 $\pm$ 0.01 $\pm$ 0.03\\
        2 & 16 & 14.80 $\pm$ 0.76  & 87.3 $\pm$ 7.8 & 0.43 $\pm$ 0.05 $\pm$ 0.03 \\
        3 & 16 & 15.71 $\pm$ 0.53  & 82.6 $\pm$ 8.8 & 0.44 $\pm$ 0.04 $\pm$ 0.03\\
        4 & 6  & 15.11 $\pm$ 0.20  & 87.7 $\pm$ 7.44 & 0.41 $\pm$ 0.01 $\pm$ 0.03\\
        5 & 16 & 15.74 $\pm$ 0.11  & 91.7 $\pm$ 10.2 & 0.43 $\pm$ 0.01 $\pm$ 0.03\\
        6 & 7  & 14.66 $\pm$ 0.11  & 92.0 $\pm$ 10.2 & 0.41 $\pm$ 0.01 $\pm$ 0.03\\
        7 & 16 & 19.27 $\pm$ 0.13  & 127.3 $\pm$ 10.1 & 0.55 $\pm$ 0.01 $\pm$ 0.03\\
        8 & 16 & 18.33 $\pm$ 0.10  & 116.8 $\pm$ 13.0 & 0.56 $\pm$ 0.01 $\pm$ 0.03\\
        \bottomrule
    \end{tabular}
    \label{tab:flux_ds_tab}
\end{table*}

\begin{table*}[htpb]
    \centering
    \renewcommand{\arraystretch}{1.3}
    \footnotesize
    \caption{Average $\gamma$-ray fluxes of the most prominent emission lines observed across the 57--2800~keV energy range. The reported values correspond to the average flux measured at the eight Hall~C positions around DS--20k. The first uncertainty represents the RMS spread among positions, while the second accounts for the systematic uncertainty.}
    \setlength{\tabcolsep}{9pt}
    
 \begin{tabular}{ccc}
        \toprule
        
        \textbf{E (keV)} & \textbf{Nuclide}  & \textbf{Flux ($\times10^{-4}\textbf{ }\mathrm{cm^{-2}\,s^{-1}}$)} \\
        \midrule
        238.6 & $^{212}$Pb & 3.3 $\pm$ 2.1 $\pm$ 0.2\\
        351.9 & $^{214}$Pb & 13.3 $\pm$ 2.7 $\pm$ 0.8\\    
        583.2 & $^{208}$Tl & 4.2 $\pm$ 1.9 $\pm$ 0.3\\    
        609.3 & $^{214}$Bi & 21.4 $\pm$ 3.6 $\pm$ 1.3\\   
        768.4 & $^{214}$Bi & 2.6 $\pm$ 0.4 $\pm$ 0.2\\    
        911.2 & $^{228}$Ac & 4.8 $\pm$ 1.8 $\pm$ 0.3\\    
        1120.3 & $^{214}$Bi & 9.8 $\pm$ 1.7 $\pm$ 0.6\\
        1238.1 & $^{214}$Bi & 4.1 $\pm$ 0.5 $\pm$ 0.3\\
        1460.8 & $^{40}$K & 29.4 $\pm$ 8.2 $\pm$ 1.8\\        
        1764.5 & $^{214}$Bi & 12.5 $\pm$ 1.6 $\pm$ 0.8\\
        2204.2 & $^{214}$Bi & 4.4 $\pm$ 0.5 $\pm$ 0.3\\
        2614.5 & $^{208}$Tl & 10.7 $\pm$ 3.5 $\pm$ 0.7\\
        
        \bottomrule
    \end{tabular}
    \label{tab:flux_ds_tab_isotopes}
\end{table*}

\section{Study of the correlation between environmental \texorpdfstring{$\gamma$}{gamma}-rays and Rn activity at position~8}
\label{sec:radon}

It is well known that radon gas emanates naturally from the rock surrounding underground laboratories. In particular, $^{222}$Rn, an intermediate product of the $^{238}$U decay chain, has a half--life of 3.8~days and can persist in ambient air for several days. As a noble gas, Rn can be transported away from rock surfaces by air circulation and contributes to the environmental radiation mainly through its short-lived decay products, $^{214}$Pb and $^{214}$Bi. In the experimental hall, significant temporal variations of the radon activity can occur, including periods in which radon peaks reach several hundred Bq/m$^{3}$.

To study the impact of the ambient Rn activity on the measured $\gamma$--ray radiation, the measuring station was equipped with a portable $\alpha$--spectrometer, as described in Sec.~\ref{sec:Setup}. The $\gamma$--ray flux at the different measurement positions was determined from data sets with typical durations of a few hours (all shorter than one day, see Table~\ref{tab:flux_ds_tab}), which were sufficient to achieve the required statistical precision and reduced the sensitivity to radon activity variations occurring on longer time scales.

\begin{figure*}[htpb]
\centering
\includegraphics[scale=0.31]{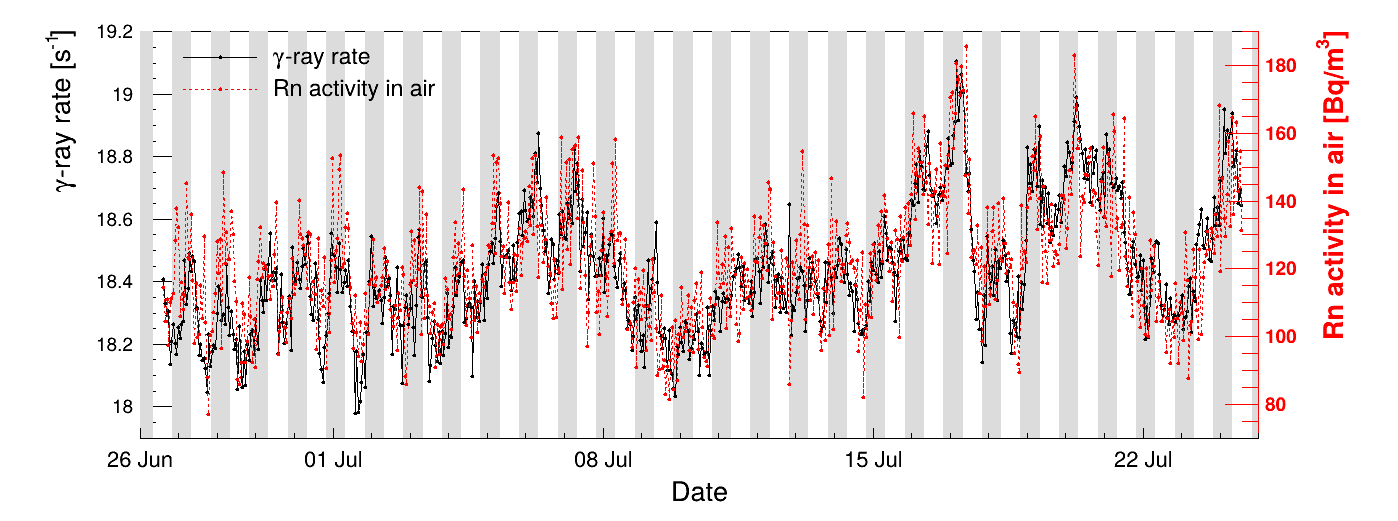}
\caption{$\gamma$--ray rate (black, left axis) and radon activity in air (red, right axis) in Hall~C as a function of time, from 26~June to 23~July~2025. Each data point corresponds to a 1~h measurement interval. White and gray bands indicate daytime (8~a.m.--8~p.m.) and nighttime (8~p.m.--8~a.m.) periods, respectively.}
\label{fig:rn_rate}
\end{figure*}

\begin{figure}[htpb]
\centering

\includegraphics[width=0.51\textwidth]{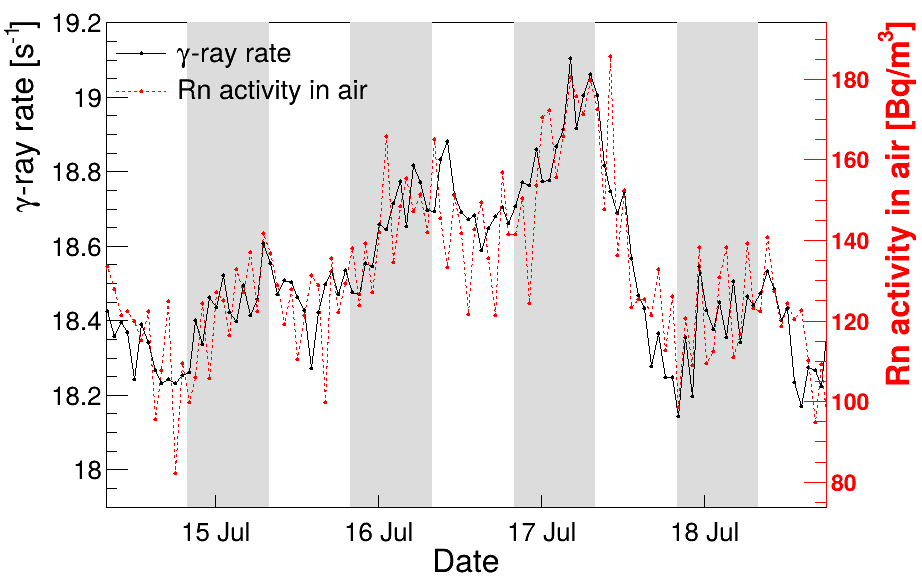}
\hfill
\includegraphics[width=0.47\textwidth]{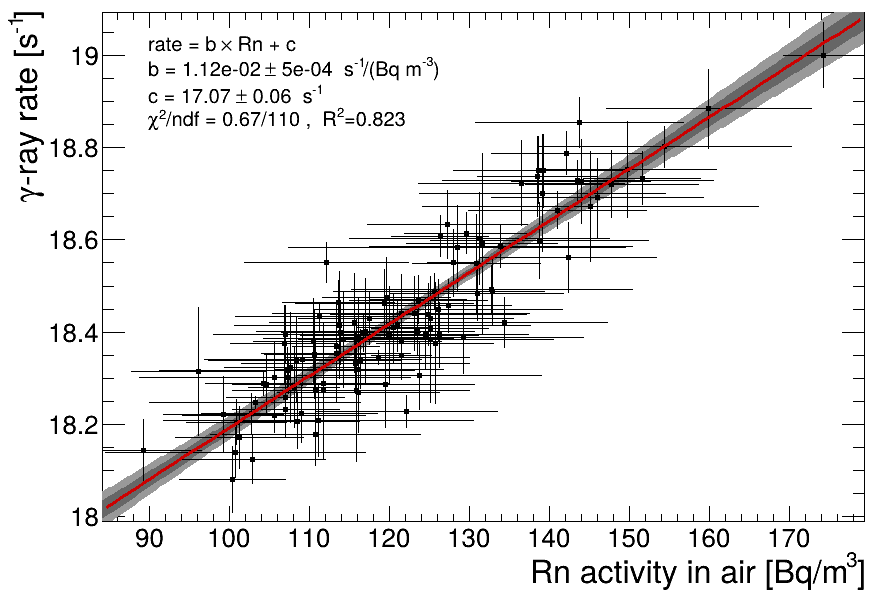}

\caption{Left: Time evolution of the measured $\gamma$--ray rate and radon activity (1~h time intervals) in air in Hall~C over a four--day period. The shaded bands indicate nighttime intervals. 
Right: Correlation between the binned mean $\gamma$--ray rate and the binned mean radon activity in air in Hall~C (6~h time bins). The red line shows a linear fit. The shaded bands represent the fit confidence intervals for the mean response (inner band: 68\% confidence level; outer band: 95\% confidence level).}

\label{fig:rn_combined}

\end{figure}

In order to investigate the correlation between the $\gamma$--ray rate and the radon activity, an additional dedicated long--term run, distinct from those reported in Table~\ref{tab:rates_ds_tab}, was carried out at position~8. Figure~\ref{fig:rn_rate} shows the time evolution of the average $\gamma$--ray rate measured by the HPGe detector and the radon activity in air in Hall~C over an extended period, from 26~June to 23~July~2025. A clear time dependence of the $\gamma$--ray rate is observed, exhibiting a pronounced temporal modulation that closely follows the day--to--day variations of the radon activity. This behavior demonstrates that the environmental $\gamma$--ray field in Hall~C is not stationary, but varies with time depending on the radon activity present at the moment of data taking.
The measured rate can be interpreted as the superposition of a radon--dependent contribution, driven by the activity of short--lived radon progeny in air, and a radon--independent baseline originating from other environmental and structural sources in the hall. Superimposed on this global radon dependence, faster short--term ($\approx$ 1~h) fluctuations are also visible, likely related to local and transient effects. The overall evolution of the $\gamma$--ray rate, however, is largely governed by the radon activity in air. This interpretation is consistent with the contribution of the prominent $\gamma$--ray lines emitted by short--lived radon daughters such as $^{214}$Pb and $^{214}$Bi.
To highlight possible day--night variations, the white and gray bands in Fig.~\ref{fig:rn_rate} indicate daytime (8~a.m.--8~p.m.) and nighttime (8~p.m.--8~a.m.) periods, respectively. The data show that radon activity---and consequently the $\gamma$--ray rate---tend to increase during the night (Fig.~\ref{fig:rn_combined}--left).

This behavior can be attributed to the ventilation conditions in the hall: during daytime, regular laboratory activity and door openings promote air exchange, while at night reduced air circulation favors radon accumulation within the enclosed volume.

To quantify the dependence of the $\gamma$--ray rate on the radon activity while reducing the impact of the faster ($\approx$ 1~h) fluctuations discussed above, the data were averaged in 6~h time bins. The binned mean $\gamma$--ray rate was then plotted as a function of the corresponding binned mean radon activity in air. The corresponding RMS spreads within each bin are shown as error bars.

Figure~\ref{fig:rn_combined}--right shows the resulting correlation scatter plot together with a linear fit, highlighting a clear linear trend ($R^{2}=0.823$). The fitted slope quantifies the sensitivity of the $\gamma$--ray rate to variations in the ambient radon activity, such that an increase of 1 Bq\,m$^{-3}$ in the average radon activity corresponds to an increase of approximately 0.011~Hz in the measured $\gamma$--ray rate. The intercept defines the radon--independent baseline rate at position~8 in Hall C. The fit yields a value of $17.07 \pm 0.06$~s$^{-1}$ for this baseline, to which the radon--induced component adds linearly.

For typical radon activity levels observed in Hall C (of order 100~Bq\,m$^{-3}$), the radon--induced contribution amounts to approximately 1.1~Hz. This corresponds to roughly 6.5\% of the total baseline rate, indicating that the dominant component of the environmental $\gamma$--ray field originates from the surrounding rock and laboratory structures, while radon provides a smaller, though measurable, additive contribution.

These results indicate that temporal variations of the $\gamma$--ray rate are mainly driven by changes in radon activity, whereas the intrinsic baseline component remains the dominant contribution to the absolute rate. Further dedicated measurements under controlled low--radon conditions are ongoing in order to isolate the baseline component with higher precision and to refine the quantitative modeling of the radon--induced contribution. These studies will be extended to other locations within LNGS, with the goal of producing a comprehensive map of the environmental $\gamma$--ray rate throughout the laboratory.

\section{Conclusions}
\label{sec:Conclusions}

Precise knowledge of the environmental $\gamma$--ray flux is a key requirement for low--background experiments operating in underground laboratories, as it directly impacts background modeling, shielding design, and sensitivity projections. 

Using a self--contained and movable HPGe detection system, we performed the first systematic spatial mapping of the environmental $\gamma$--ray flux in Hall~C. The detector response was calibrated and modeled through detailed Monte Carlo simulations validated with calibrated $\gamma$--ray sources, with particular attention devoted to the accuracy of the detector geometry description and to the treatment of systematic uncertainties. This approach allowed us to derive a robust, energy--dependent detection efficiency and to determine the $\gamma$--ray flux in the energy range 57--2800~keV. The resulting average flux ($0.46 \pm 0.06_{\mathrm{stat}} \pm 0.03_{\mathrm{syst}}$)~cm$^{-2}$~s$^{-1}$, provides the first precise reference value for Hall~C and reveals non--negligible position--dependent flux variations within the hall.

The setup enabled the simultaneous monitoring of the environmental $\gamma$--ray rate and the ambient radon activity at the same location. A clear temporal correspondence between the $\gamma$--ray rate and the radon activity in air was observed over an extended data--taking period of five weeks, including coherent day--night variations. The observed rate modulation is consistent with the contribution of short--lived radon progeny, whose decays produce prominent $\gamma$--ray lines in the measured spectrum. These findings demonstrate that the environmental $\gamma$--ray flux in Hall~C is not only position--dependent but also time--dependent. The reported flux values should therefore be understood as corresponding to the radon activity present in air during each individual measurement.

The variations of the $\gamma$--ray rate on time scales of a few hours are largely accounted for by changes in the radon activity in air. This behavior indicates that the measured rate can be expressed as the sum of a radon--dependent contribution and a constant baseline component. The extrapolation to zero radon activity identifies a non--zero baseline component associated with the intrinsic $\gamma$--ray emission from the surrounding rock and structural materials of the hall. For typical radon activity levels observed in Hall~C (of order 100~Bq\,m$^{-3}$), the radon--induced contribution amounts to approximately 1.1~Hz. This corresponds to about 6--7\% of the baseline rate.

Overall, this work provides a comprehensive and quantitatively controlled characterization of the radiological environment in Hall~C, supplying essential input for current and future rare--event searches hosted in this hall, such as DarkSide-20k and CUPID. The observed interplay between $\gamma$--ray radiation and radon activity highlights the importance of joint and continuous monitoring of these environmental parameters. 

Within the framework of the ongoing collaboration between CIEMAT, GSSI, LNGS, and Jagiellonian
University, new measurements of the $\gamma$--ray rate and radon activity are currently in progress, with the objective of achieving a detailed and systematic mapping of the environmental $\gamma$--ray field throughout the entire LNGS laboratory. 

The mobile nature of the measurement station further makes it suitable for deployment in other underground laboratories, should there be interest in performing comparable measurements and establishing a consistent comparison of the environmental $\gamma$--ray flux using the same calibrated hardware.

\section*{Acknowledgments}
\label{sec:Acknow}

The authors would like to thank Dr. Davide Franco (APC) for his collaboration and support in the Monte Carlo simulations.
They also gratefully acknowledge the LNGS Director and the LNGS Services for their continuous support and assistance throughout the measurements.

This work was made possible by funding from the Spanish Ministry of Science and Innovation (MICINN) under Grant PID2022--138357NB--C22. It also received support from the Polish National Science Centre (NCN) through Grant No. UMO-2023/50/A/ST2/00651. Research for this publication was funded through the budget of the Anthropocene Priority Research Area (Earth System Science Core Facility Flagship Project) under the Strategic Programme Excellence Initiative at the Jagiellonian University.

\bibliographystyle{apsrev4-2}

\begin{thebibliography}{99}

\bibitem{haffke}
M. Haffke et al., Background Measurements in the Gran Sasso Underground Laboratory, Nuclear Instruments and Methods in
Physics Research A, 643, 36-41 (2011), doi: \href{https://doi.org/10.1016/j.nima.2011.04.029}{10.1016/j.nima.2011.04.029}.

\bibitem{malceski}
Dariusz Malczewski et al., Gamma background measurements in the Gran Sasso National Laboratory, J Radioanal Nucl Chem, 295, 749–754 (2013), doi: \href{https://doi.org/10.1007/s10967-012-1990-9}{10.1007/s10967-012-1990-9}.
 
\bibitem{bucci}
C. Bucci et al., Background study and Monte Carlo simulations for large-mass bolometers, Eur. Phys. J. A 41, 155–168 (2009), doi: \href{https://doi.org/10.1140/epja/i2009-10805-7}{10.1140/epja/i2009-10805-7}.  

\bibitem{arpesella}
C. Arpesella, Background measurements at Gran Sasso Laboratory, Nucl. Phys. B (Proc. Suppl.) 28A, 420-424 (1992), doi: \href{https://doi.org/10.1016/0920-5632(92)90207-9}{10.1016/0920-5632(92)90207-9}.

\bibitem{ds-20k}

The DarkSide-20k Collaboration, DarkSide-20k sensitivity to light dark matter particles, Communications Physics, 7, 422 (2024) , doi: \href{https://doi.org/10.1038/s42005-024-01896-z}{10.1038/s42005-024-01896-z}.

\bibitem{cupid}

The CUPID Collaboration, CUPID, the Cuore upgrade with particle identification, Eur. Phys. J. C (2025), 85, 7, 737, note = "[Erratum: Eur.Phys.J.C 85, 1346 (2025)]", doi: \href{https://doi.org/10.1140/epjc/s10052-025-14352-1}{10.1140/epjc/s10052-025-14352-1}.

\bibitem{nomad}
Evaluating the Commercial Spectrometer Systems for Safeguards Applications Using Germanium Detectors, D.T.Vo, Los Alamos National Laboratory, NM 87545, available online at \href{https://www.ortec-online.com/-/media/ametekortec/technical-papers/digital-gamma-ray-spectroscopy-systems/evaluating-commercial-spectrometer-systems-safeguards-applications-using-germanium-detectors.pdf}{ORTEC technical report}.

\bibitem{alpha}
Bertin Technologies, User Manual - AlphaGUARD (2019), available online at \href{https://www.laurussystems.com/wp-content/uploads/USER-MANUAL-ALPHAGUARD-2019.pdf}{Bertin Technologies AlphaGUARD}.

\bibitem{radionuclide}
Wolfgang Wahl, Radionuclide Handbook for Laboratory Workers in Spectrometry, Radiation Protection and Medice, Institute for Spectrometry and Radiation Protection, Version 5.3.1 (2016), catalogued by the United Nations Vienna Library:
\url{https://unov.tind.io/record/69733}

\bibitem{Knoll}
G. F. Knoll, Radiation Detection and Measurement, 4th ed. (John Wiley and Sons, Hoboken, NJ, 2010), available online at \url{https://indico-tdli.sjtu.edu.cn/event/171/contributions/2123/attachments/982/1592/Knoll4thEdition.pdf}

\bibitem{lutz}
G. Lutz, Semiconductor Radiation Detectors: Device Physics, Springer, Berlin (1999), available online at \url{https://link.springer.com/book/10.1007/978-3-540-71679-2}.

\bibitem{hafiz}
N. Hafızoğlu, Efficiency and energy resolution of gamma spectrometry system with HPGe detector depending on variable source-to-detector distances, Eur. Phys. J. Plus, 134 (2024), doi: \href{https://doi.org/10.1140/epjp/s13360-024-04903-y}{10.1140/epjp/s13360-024-04903-y}.

\bibitem{geant}
[Online]. Available at \url{https://geant4.web.cern.ch/}.

\bibitem{lee}
Hyeonmin Lee et al., Dead layer estimation of an HPGe detector using MCNP6 and Geant4, Applied Radiation and Isotopes, 192 (2023), doi: \href{https://doi.org/10.1016/j.apradiso.2022.110597}{10.1016/j.apradiso.2022.110597}. 

\bibitem{jesko}
M. Ješkovský et al., Experimental and Monte Carlo determination of HPGe detector efficiency, Journal of Radioanalytical and Nuclear Chemistry 322:1863–1869 (2019), doi: \href{https://doi.org/10.1007/s10967-019-06856-4}{10.1007/s10967-019-06856-4}.

\bibitem{jany}
A. Jany et al., Fabrication, characterization and analysis of a prototype high purity germanium detector for ${}^{76}\text{Ge}$-based neutrinoless double beta decay experiments, The European Physical Journal C, 81 38 (2021), doi: \href{https://doi.org/10.1140/epjc/s10052-020-08781-3}{10.1140/epjc/s10052-020-08781-3}.


\bibitem{john}
J. Greenwood, The correct and incorrect generation of a cosine distribution of scattered particles for Monte-Carlo modeling of vacuum systems, Vacuum 67, 217--222 (2002), doi: \href{https://doi.org/10.1016/S0042-207X(02)00173-2}{10.1016/S0042-207X(02)00173-2}.


\end{thebibliography}

\end{document}